# Implementation of Recommendation Algorithm based on Recommendation Sessions in E-commerce IT System

Michał Malinowski

Military University of Technology, Faculty of Cybernetics, Poland

**ABSTRACT**

The aim of the paper is to present a study as the implementation of the author's Algorithm of the Recommendation Sessions ARS in an operating e-commerce information system and to analyse basic parameters of the recommendation system created as a result of the implementation. The first part of the study contains a synthetic description of the area of recommendation systems. The next section presents the proprietary ARS recommendation algorithm based on recommendation sessions. The third part of the paper describes the mathematical model of the recommendation session built on the basis of the theory of graphs and networks, which such model makes the input data for the algorithm in question. The next part of the publication describes the possibilities of representing graph structures and the method of implementing a G graph (constituting a set of the recommendation session) in a relational database. The implementation of the ARS algorithm, based on the SQL standard, was also presented. The implementations in question have been developed on the basis of a working information system of the e-commerce class. As a result of the implementation of the algorithm, a fully functional recommendation system was created, which can be adapted to various e-commerce IT systems. The positive result of the work was confirmed by the research on the parameters of the recommendation system, included in the last part of the study.

## Introduction

There is no doubt that recommendation solutions are becoming more and more important in the modern consumer economy. Such systems are used, among others, in large web portals, search engines, social networks and online shops. The information environment we live in, and in particular the huge amount of information provided to us, makes it impossible to use it effectively (Malinowski &Sokólski, 2001). For this purpose, dedicated algorithms are created, and hence IT systems based on them, the main function of which is to support the decision-making process.

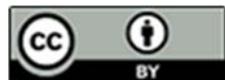

**Related works**

The recommendation area is considered under three main domains (Liling, 2019):
- recommendation techniques –in the context of the development of analytical methods of the recommendation problem;
- recommendation algorithms –in the context of developing algorithms based on recommendation techniques;
- recommendation systems–in the context of building IT solutions that support recommendation processes using recommendation algorithms.

The recommendation systems, also often referred to as "recommender systems", are used to estimate users' preferences for items, services or objects which they have not yet seen or are not familiar with. Recommendation systems often use inputs such as users' preferences, features (attributes) of an item (object), history of users' past interactions with items (objects), temporal data, and spatial data (Bhaskar& Raja, 2020).

Recommendations provided by these types of systems come in two different forms:
- prediction of ratings – in this case, the user's rating for a new item is estimated. The prediction is based on the ratings of the items considered by other users for the set. For example, a prediction of a rating is a forecast of whether or not a particular user will like a certain movie on Netflix, based on the historical behaviour of other Netflix users (Symeonidis et al., 2011);
- ranking forecast – a result estimation involves creating a "Top–N" ranking list with N positions for a particular user. A system based on this schema can recommend "Top 10 Books To Read If You Like Harry Potter" (Cremonesi et al., 2010).

In e-commerce solutions, recommendation systems that produce results as ranking forecasts rather than in the form of predicted ratings are preferred because companies want to have a list of expected products rather than ratings of a set of products (Steck, 2013).

The most important feature of the recommendation system is its ability to predict the user's preferences and interests by analysing his behaviour or the behaviours of other users in order to generate personalised results.

Currently, the following main types of techniques are distinguished in the studies concerning the recommendation area (Ricci et al., 2010, p. 10):
- Content-based Methods;
- Collaborative Filtering Methods;
- Memory-based;
- Model-based;
- Hybrid Methods.

**Content-based methods**
Algorithms based on this technique are constructed on the assumption that objects with similar functions will receive similar ratings from the same user. Consequently, systems of this type recommend objects that are similar to items about which the user showed his interest in the past. The similarity of objects is calculated based on specific attributes and using various methods (Li et al., 2012).

**Collaborative filtering methods**
The idea behind this technique is based on the assumption that if two users have made similar decisions in the past, their preferences will also coincide in the future. An example would be a board game store where two users showed an interest in similar games, but one of them was interested in other games as well. In such a situation, the common filtration mechanism will also recommend these games to the other user, assuming that he will also like them. In the above situation, the system does not base its operation on the attributes of objects or users, but only on the behaviour of people whose interests and preferences have been found to be similar (Su&Khoshgoftaar, 2009).

**Memory-based**
These methods are characterised by the fact that they use the entire database of the website to carry out recommendations. They assume that each user is part of a certain group; therefore, they use statistical methods to find the so-called "closest neighbours", that is, groups of users with similar interests. In the first step, such recommendation mechanism searches for a group of the closest neighbours by calculating the weight in $_{i,j}$, which determines the similarity or correlation between two users: *i* and *j*. Next, the forecast determining whether a given product can be recommended to the user based on its closest neighbours is calculated. The last step is to select N objects most suitable for the user and present them as recommendations (Su&Khoshgoftaar, 2009).

**Model-based**
The basis for their operations is the creation of a user rating model that can predict the user's rating of objects. Machine learning techniques based on training data are used to create the model (Sarwar et al., 2001).One way to estimate the recommendation of an object for a given user is to use the Bayesian classifier, which is most commonly used in recommendation systems. It is one of the machine learning methods that determines to which of the decision classes a new case should be assigned. In the context of recommendation mechanisms, it is determined how suitable the selected object is for a given user (Miyahara &Pazzani, 2000).

**Hybrid methods**
Such techniques use the properties of both previously described approaches and compile both Content-based Filtering Methods and Collaborative Filtering (CF) Based Methods. Hybrid Methods overcome the disadvantages of individual approaches.

Hybrid techniques heavily use users' profiles and objects' descriptions to find users with similar interests, then use collaborative filtering to make predictions (Cremonesi et al., 2010).

**The research problem**
The recommendation algorithms are designed based on the techniques that operate and are advanced in the sphere of ideas. As a result, different algorithms can be developed based on the same techniques and can produce the same results. The times of generating the results and the size of the required resources (processor power and memory) make significant differences between them are (Goodrich &Tamassia, 2001, "2.2" section). Ultimately, the algorithms are implemented with the use of computer techniques in recommendation systems, which are used as part of e-commerce solutions.

In general, the task of recommendation algorithms is to solve the recommendation task described as follows (Esmaili et al., 2006). Let $C$ be a set of users using the recommendation system and $O$ a set of all possible objects that can be recommended. In addition, let $u$ be a utility function measuring the utility of object $o \in O$ for a user $c \in C$. This is $u: CxO \rightarrow W$, where $W$ is an ordered set (e.g. non-negative integers) Then, for each user $c$ belonging to $C$, a sub-set $R_c^*$ belonging to $O$ is selected, which maximises the usability for the user. This means that for the arranged $c \in C$ we determine the set of the recommended objects as follows:

$$R_c^* = \{r \in O: u(c,r) = \max_{o \in O} u(c,o)\} \quad (1)$$

In a general case, the task of the recommendation algorithms is to find a sub-set $R_c^*$ called the set of recommended objects for a user **c**.

There are two main problems when building such algorithms: handling sparse user data (Shams & Haratizadeh, 2017) and the need for implicitly derived information (McAuley&Leskovec, 2013). In connection with the above, the task is modified as follows: the recommendation is made for the selected object $m$ without directly taking into account the user $c$, for whom the recommendation is made. In practice, such an event occurs when the recommendation system does not have information about the user. This may occur in the following situations: the user is exposed to the system for the first time, or the system does not collect information about the users.

Therefore, task (1), taking into account the change in the utility function such that $u: OxO \rightarrow W$, is subject to change so that the set of recommendations for the user $R_c^*$ is replaced by $R_m^*$, i.e. the set of recommendations for the object. Consequently, for the arranged $m \in O$, we determine the set of recommended objects as follows:

$$R_m^* = \{r \in O: u(m,r) = \max_{o \in O} u(m,o)\} \quad (2)$$

Considering user $c$ consists in the fact that it is him who, consciously directing his behaviour, selects an object $m$ from the $O$ set, against which recommendation $R_m^*$ shall be built up.

**The solution to the problem**
In order to solve the presented special task (2),the Recommendation Algorithm Based On Recommendation Sessions was developed (ARS) (Malinowski, 2020). The algorithm can be classified as the hybrid algorithm type because it can make recommendations based on both the Content-based Method and the Collaborative Filtering Method.

The solution proposes to build a data model based on the concept of the recommendation session. In the area of mathematical apparatus, this concept is defined on the basis of a bipartite directed unigraph **G**(Wojciechowski, 2019), called a recommendation session graph, such that:

$$G = <N, E> \quad (3)$$

where:
$N = J \cup O$ – the set of vertices (node)
$E \subset J \times O$ – the set of arcs (directed edges)
$O$ – a set of objects where **an object** ($o \in O$) may be:
- commodity in an Internet shop;
- a video in a video-on-demand service;
- an employee in an employment-related service;
- press article in the news service;
- a person on a social network application.

$J$ – a set of kernels, where **the kernel**($j \in J$) may be:
- a products' category – one of the sub-sets of products having common features;
- the purchase order – the result of the customer's actions in the shop, which is finalised with the purchase;
- a wish list – a sub-set of the shop's products related to the customer, which such sub-set results from the customer's future shopping preferences;
- an expert – a sub-set of products indicated by the discipline specialist;
- a website visit ID – a unique key assigned to a user's visits to the website. A visit consists of a sequence of viewed/visited pages of an online store;
- a person – IT structure identifying and describing the user in the IT system.

Due to the adopted and further discussed the interpretation of the model and its components, the following constraints are assumed to be met:
- $J \cap O = \emptyset$ – no kernel shall be a facility, and no facility shall be a kernel
- $(\forall j \in J)(\exists o \in O)(\exists e \in E)(e = (j, o))$ – each kernel must be associated with at least one object
- $(\forall o \in O)(\exists j \in J)(\exists e \in E)(e = (j, o))$ – each object must be associated with at least one kernel

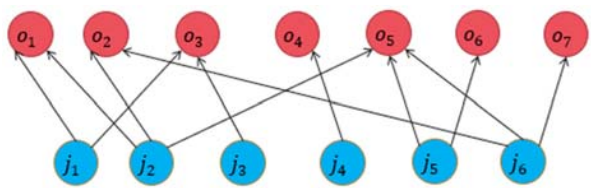

*Figure 1.*The recommendation session's graph

Note.objects $\{o_1, o_2, o_3 \ldots o_n\}$, kernels $\{j_1, j_2, j_3 \ldots j_n\}$

A single session **S** can be presented as a sub-graph of the graph **G** such that:
$$S = <N', E'> \qquad (4)$$
where:
$N' = j \cup O'$ – the set of session's vertices
$E' \subset E$ – the set of the session's arcs
$O' \subset O$ – a set of session's objects (related to a kernel)
$j \in J$ – the session's kernel
It is assumed that the following constraints are met:
- $(\forall o \in O')(\exists e \in E)(e = (j, o))$ – each object of a session is associated with a kernel
- $(\nexists o \in O \setminus O')(\exists e \in E)(e = (j, o))$ – there is no kernel-associated object that is not a part of sessions

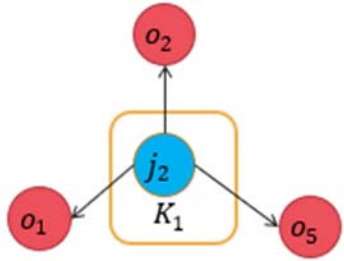

*Figure 2.* Single recommendation session model.
Note. Objects $\{o_1, o_2, o_3\}$, kernel $\{j_2\}$ and class $\{K_1\}$

The session is made up of one kernel only and of all arcs and objects related to the kernel.
S – set of sessions, where **the session** ($s \in S$) may be:
- the products' category and the products themselves;
- the purchase order and the order's items;
- the customer's wish list with the elements;
- an expert with his/her assessments;
- an identifier (ID) of the website visited, along with the pages visited;
- a person with his/her colleagues.

Moreover, because of its physical properties and similarities, let us denote by **K** the set of different types of the session's kernels. Each $k \in K$ corresponds to $J_k \subset J$ a subset of kernels of the same type, in other words, with the same functional properties, e.g., product categories, orders, etc.
Which means:
$K \subset J$ – the class is the sub-set of the kernels' set
Where **the class** ($K \subset J$) of the session's kernels may be:
- the categories of the products;
- the purchase orders;
- the customers' wish lists;

- experts;
- identifiers (IDs) of the website visits;
- persons.

In addition, the following constraints are assumed to be met:
- $(\forall i \neq j)(K_i \cap K_j = \emptyset)$ – no kernel is present in two or more classes
- $(\forall j \in J)(\exists i)(j \in K_i)$ – each kernel is a member of a class

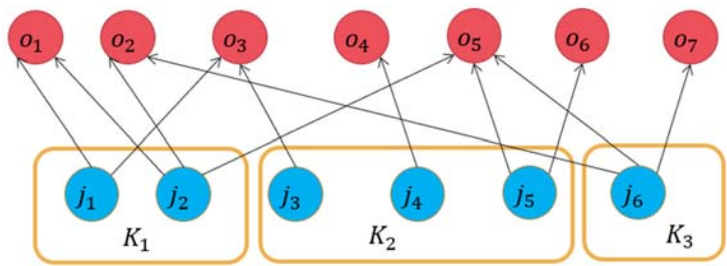

*Figure 3.* The classes of the kernels

Note. Objects $\{o_1, o_2, o_3 \ldots o_n\}$, kernels $\{j_1, j_2, j_3 \ldots j_n\}$ and classes $\{K_1, K_2, K_3\}$

Generally, the classes of kernels can be divided into the following types:
- behavioural – kernels created as a result of users' actions. They are highly variable over time, such as visiting, shopping or wish list;
- static – kernels related to the features of objects. They are invariable over time, such as size, colour or category membership;
- mixed – kernels formed as a result of the influence of the system environment, but poorly variable over time, such as the indications of experts or external rankings.

The ARS algorithm consists of specific types of input data, output data and steps: They are discussed in detail with examples in the paper *"Recommendation Algorithm Based on Recommendation Sessions"* included in the conference materials of the "36th International Business Information Management Association (IBIMA)" (Malinowski, 2020). In this study, only the basic elements of the algorithm will be presented, such as:

*Input data*
**G** – the session's graph
**m** – the object (graph node) to which the recommendations are to be linked

*Output data*
**R$_m$** – vector of recommendation for the object **m** (the lower the position in the vector, the better the recommendation for the object **m**)

Formally, the relationship between the set $\boldsymbol{R}_m^*$ (2) and the recommendation vector $\boldsymbol{R_m}$ is such that each element of the recommendation set $\boldsymbol{R}_m^*$ is the element of the recommendation vector, but not every element of the vector $\boldsymbol{R_m}$ is the element of the set $\boldsymbol{R}_m^*$. This is due to the fact that the algorithm returns not only the best match but also results close to the best N objects.

*Steps:*
*S01:* Presentation of input data
*S02:* The construction of the sub-graph **G'<sub>m</sub>** of the graph **G** consisting of node **m** and nodes adjacent to the node **m** and arcs between them and node **m**
*S03:* The construction of the sub-graph **G''<sub>m</sub>** of the graph **G** composed of the graph **G'<sub>m</sub>** and the nodes adjacent to the nodes of the graph **G'<sub>m</sub>**, and arcs between them and the nodes of the graph **G'<sub>m</sub>**
*S04:* Estimation of the incoming steps for each node that is the object of sub-graph **G''<sub>m</sub>**
*S05:* Descending sorting of objects (nodes) in relation to the incoming step
*S06:* Saving sorted objects in the vector **R<sub>m</sub>**, without the object **m**

**Implementation of the problem**
The main problems with the implementation of the algorithm in the recommendation system are the construction of the graph **G** in the database structures of the target e-commerce IT system and the implementation of the algorithm steps using the IT mechanisms of a given e-commerce solution.

**The construction of the graph G**
Graphs are an abstract structure that does not have its direct reflection in the environment. Their elements, namely vertices and edges, can have differing quantities. Moreover, the mathematical model of graphs is not naturally reflected in the memory organisation of computers (Horzyk, 2018). When implementing graphs, other well-defined programming structures such as arrays and lists are used. They are generally used to store information about neighbouring (incidental) vertices or edges connecting them (Goodrich &Tamassia, 2001, Chapter 6). Graphs are usually represented and implemented in the form (Cormen, 2009, Chapter 6) of:

- adjacency matrix –a matrix is presented as a two-dimensional array, where the row and column indexes represent vertices' numbers, and elements' values equal to 1 mean the edge connecting the vertices denoted by the row number and the column number;
- incidence matrix –a matrix is presented as a two-dimensional array, where row indexes represent vertices' numbers, and column indexes represent edges' numbers. Elements equal to 1 represent edges denoted by numbers of incident columns with vertices denoted by numbers of rows;
- adjacency list –a list is represented as the list where indexes represent vertices' numbers, and each element of this list is a list of numbers of neighbouring vertices with the vertex number being present in the index;
- incidence list –a list is represented as the list where indexes represent numbers of vertices, and each element of this list is a list of numbers of incident edges with the vertex number being present in the index;
- list of edges –a list is represented as the list of pairs of vertices' numbers for each edge.

22

Management and Business Research Quarterly 2021(19)14–32

Table 1.Comparison of graph representations in computers' memory

| College | Complexity of memory |
|---|---|
| Adjacency matrix | $O(|N|^2)$ |
| Incidence matrix | $O(|N|*|E|)$ |
| adjacency list | $O(|N|+|E|)$ |
| incidence list | $O(|N|+|E|)$ |
| list of edges | $O(|E|)$ |

Note. |N| is the count of the vertices set, |E| is the count of the edge set.

A representation in the form of the edge list was chosen to represent graph **G** in the target e-commerce system.

The implementation of the ARS recommendation algorithm was carried out in the e-commerce solution in the form of an online store platform, with board games, operating at the URL address https://am76.pl, called AM76.The SQL standard implemented in the relational database of the AM76 information system was adopted as the method of implementation of the algorithm. Figure 4 shows a fragment of the database in question with the structure of data essential for the construction of graph G.

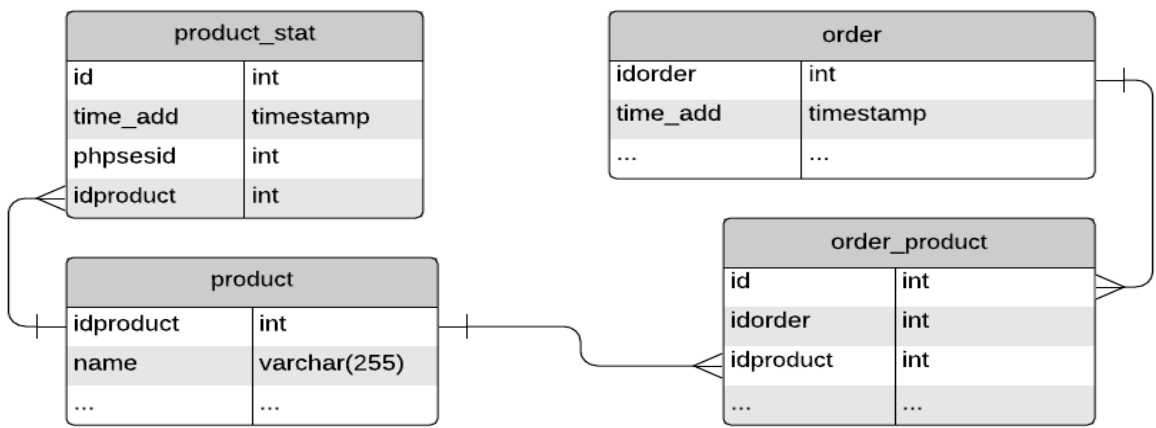

*Figure 4.* The data structure in the form of an ERD diagram of the AM76 information system database
Note. The tables "product", "product_state", "order" and "order_product."

The data used to build graph $G = <N, E>$ were the data sets such as:
$N = O \cup J$ – the set of objects composed of the sum of two sets
$O$ – objects, products that are board games. Information about them is contained in the *product* table, and the unique identifier is *the idproduct* field
$J = K_1 \cup K_2$ – the set of kernels composed of the sum of two classes:
$K_1$ – purchase orders for products offered by an online store Information about them is contained in the *order_product* table, and the unique identifier is the *order_product.id* field
$K_2$ – visiting products offered by the online store Information about them is contained in the *product_state* table, and the unique identifier is the *phpsesid* field
$E = E_1 \cup E_2$ – the set of the arcs composed of the sum of two sets:
$E_1$ – purchase orders for products: *order_product.id –>idproduct*

$E_2$ – visiting products offered by the online store: *phpsesid –>idproduct*

A data structure was built within the AM76 system database, which stores, according to the representation adopted, lists of edges of the graph **G**. The *kernel_object* index of UNIQUE type has been set on this structure. This key ensured that the edges do not repeat and the condition that the graph **G** is a unigraph was met.

| graph_g | |
|---|---|
| kernel | int |
| object | int |
| class | int |

*Figure 5.*Structure of the graph representation table G in the database of the AM76 information system
Note.The table "graph_g" representing the arcs of the graph G in the *kernel (class)–>object* convention

The graph **G** is built every hour based on current data from the AM76 information system. It is created as a result of a sequence of SQL queries:

<u>Query 1</u> – removing the earlier implementation of the graph G
*TRUNCATE TABLE graph_g;*

<u>Query 2</u> – proper implementation of the graph G on the basis of data from the AM76 information system
*INSERT INTO graph _g SELECT * FROM (*
*#k1*
*SELECT order_product.idorder AS kernel, order_product.idproduct AS object, 1 AS class FROM order_product*
*INNER JOIN order ON order.idorder=order_product.idorder*
*INNER JOIN product ON product.idproduct=order_product.idproduct*
*UNION*
*#k2*
*SELECT product_stat.phpsessid AS kernel, product_stat.idproduct AS object, 2 AS class FROM product_stat*
*INNER JOIN product ON product.idproduct= product_stat.idproduct*
*) AS graph_g_view;*

<u>Query 3</u> – optimization of data of the graph G
*OPTIMIZE TABLE graph_g;*

**Implementation of the algorithm steps**

Based on the constructed representation of the graph G in the form of the *graph_g* table, the individual steps of the(ARS) algorithm (Malinowski, 2020, p. 11264), in the AM76 system, took the form:

*S01:*<u>Providing input data</u>

**m** – retrieving information on the board game to be recommended from the AM76 website

**G** – using the *graph_g* table

*S02:*<u>The construction of the sub-graph **G'$_m$** of the graph **G** consisting of node **m** and nodes adjacent to the node **m** and arcs between them and node **m**</u>

The execution of a SQL query of the form:

*SELECT kernel, object FROM graph_g WHERE object=m*

*S03:*<u>The construction of the sub-graph **G''$_m$** of the graph **G** composed of the graph **G'$_m$** and the nodes adjacent to the nodes of the graph **G'$_m$**, and arcs between them and the nodes of the graph **G'$_m$**</u>

The execution of a SQL query of the form:

*SELECT kernel, object FROM graph_g*
*WHERE kernel IN(SELECT kernel FROM graph_g WHERE object=m)*

*S04:*<u>Estimation of the incoming steps for each node that is the object of sub-graph **G''$_m$**</u>

The execution of a SQL query of the form:

*SELECT object, count(\*) AS degree_in FROM graph_g*
*WHERE kernel IN(SELECT kernel FROM graph_g WHERE object=m)*

*S05:*<u>Descending sorting of objects (nodes) in relation to the incoming step</u>

One adds, to the SQL query from K04, the SQL code of the following form

*(…) ORDER BY degree in DESC*

*S06:*<u>Saving the sorted objects in the vector **R$_m$** without object **m**</u>Returning the result of the query from steps S04+S05 to the website of the AM76 platform

It should be noted that the steps of the algorithm from S02 to S05 can be written within a single SQL query of the form:

*SELECT object, count (\*) AS degree in FROM graph_g*
*WHERE object <>m*
*AND kernel IN(SELECT kernel FROM graf_g WHERE object =m)*
*GROUP BY object*
*ORDER BY degree_in DESC*

The above SQL query is the final query and returns the result of the ARS algorithm.

**Results**

The result of the implementation of the ARS algorithm, by using the method to construct it, based on the SQL standard, within a relational database, is a fully functional recommendation system functioning within the e-commerce solution. In the case of this study, the system is the IT system of the AM76 online store available at the URL address https://am76.pl.

As part of the implementation, kernels of new classes were added above the previously described classes in the e-commerce solution. Eventually, such classes were implemented in the system:

$K_1$ – purchase orders – represents product items (objects) in orders made by customers (kernels)

$K_2$ – visits – represents the products (objects) visited from one IP during one day. Such a pair is called a session (kernels)

$K_3$ – categories – represents the assignment of products (objects) to thematic categories (kernels)

$K_4$ – series – represents the thematic merging of products (objects) into thematic series (kernels)
$K_5$ – wish lists – represents products (objects) that customers (kernels) subjectively distinguish
$K_6$ – expert indications – represents products (objects) that are indicated by experts for various reasons (kernels)

**Measurements of results**

A study of the implemented recommendation system based on the ARS algorithm was conducted between the 1st and the 17th of May 2021. The results of the research are presented below:

*The parameters of the graph G*

First, the size of the graph **G** was estimated on the last day of the graph **G** survey and changes in its size were measured during the surveyed period of time

*Table 2.* The counts of the sets of the graph G

| Set | O | J | N | E | K |
|---|---|---|---|---|---|
| Name | Objects | Kernels | Nodes | Arcs | Classes |
| The count | 2433 | 77850 | 80283 | 140170 | 6 |

Note. Figures/data as of May 17, 2021.

The graph G is the relatively large graph with a total of 220,453 elements. This number increases over time, as will be shown in the results of the following measurements.

*Table 3.* The counts and the types of the class sets

| Class | $K_1$ | $K_2$ | $K_3$ | $K_4$ | $K_5$ | $K_6$ |
|---|---|---|---|---|---|---|
| The count of the kernels | 2025 | 74227 | 91 | 261 | 1244 | 2 |
| The count of the objects (associated) | 697 | 2432 | 2424 | 1330 | 1553 | 60 |

Note. Data as of May 17, 2021, the sum of the sets of kernels in the classes represents the full set of kernels of graph **G**.

Note that the objects associated with the classes in question have differing counts. It is only by implementing the different classes that it is possible to fully cover (or tend to cover) all the items that are in the e-commerce system and consequently to indicate recommendations.

The size of the graph G is not constant over time and has changed according to the graphs shown

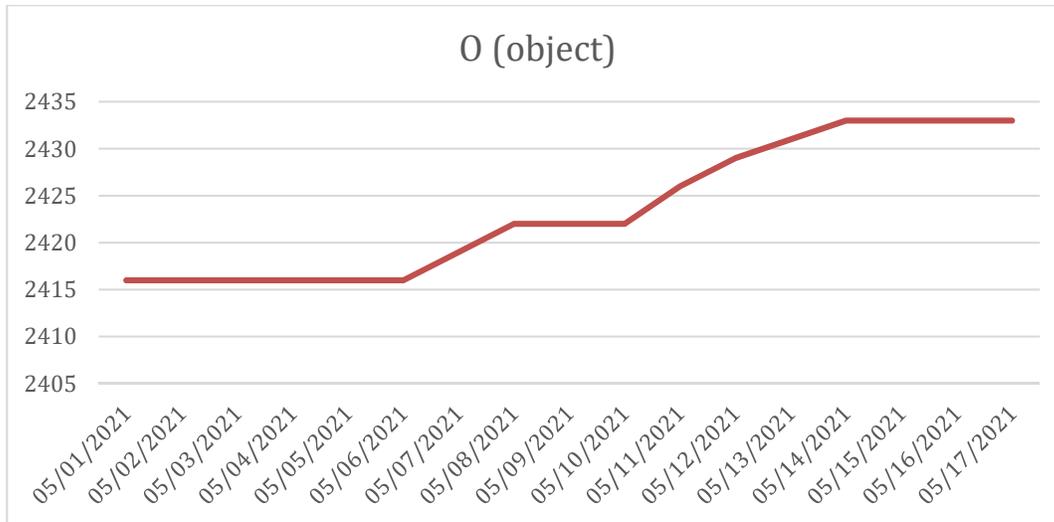

*Figure 6.* Change in count of the set O (objects) of the graph G
*Note.* The count of the set O ranged from 2416 to 2433, which means that 17 objects were added

The set O (objects) in relation to the other elements of graph G is not very numerous because it makes merely 1.10% of the number of all its elements. From the recommendation point of view, however, it is the key set because it represents the products of the e-commerce system subject to the recommendation.

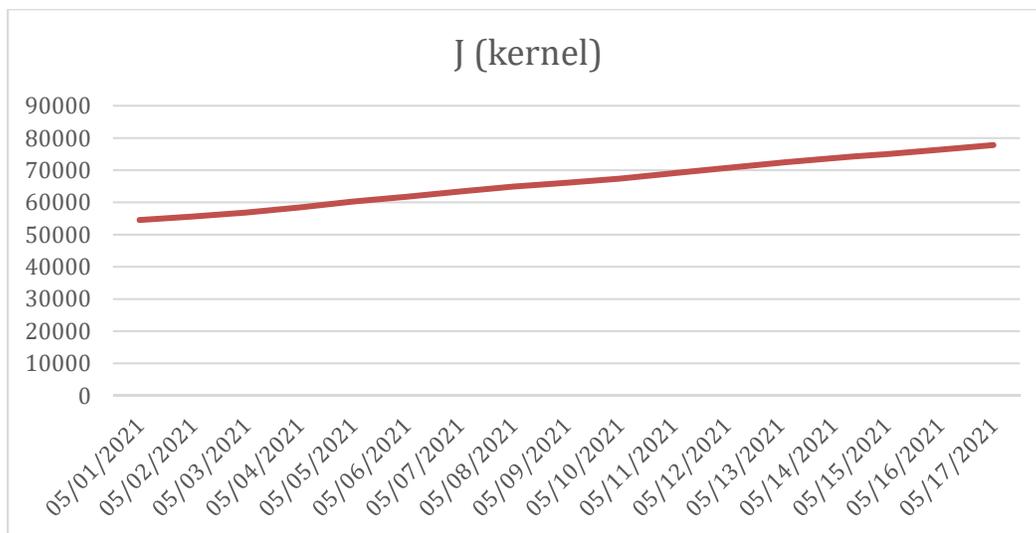

*Figure 7.* Change of the count of the set J (objects) of the graph G
*Note.* The count of the set J ranged from 54,553 to 77,850, which means that 23,297 kernels were added

The set J (kernels) in relation to the other elements of the graph G represents 35.31% of the number of all elements of the graph G and is important when constructing the elementary sessions, on which the mechanism of the ARS algorithm is based.

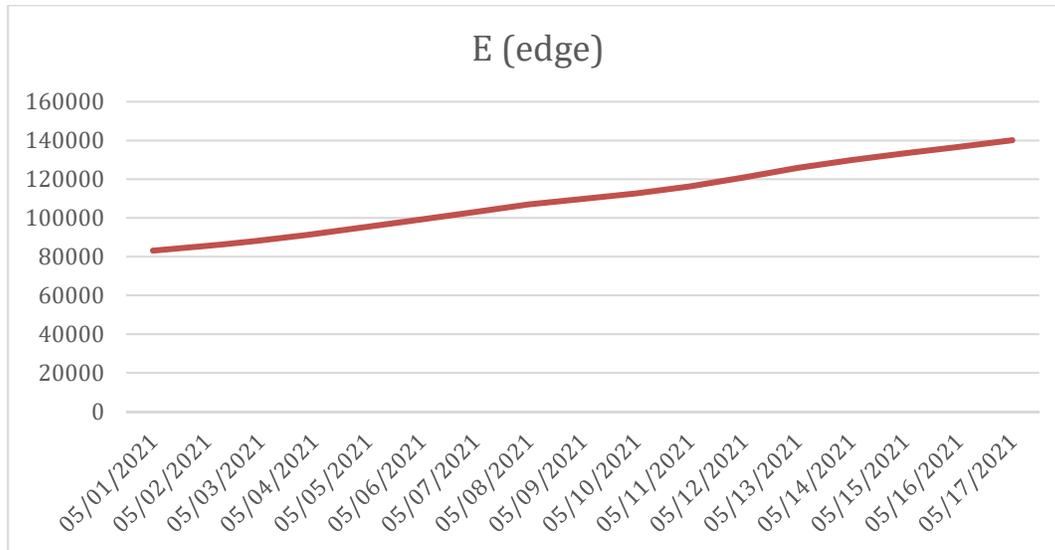

*Figure 8.* Change of the count of the set E (arcs) of the graph G
Note. The count of the set E ranged from 83,192 to 140,170, which means that 56,978 arcs were added

The set E (arcs), in relation to the other elements of graph G, represents 63.59% of the number of all elements of graph G, which is the dominant value in relation to the number of other elements. The number of these elements increases the most overtime and accounts for the quality of the results of the ARS algorithm.

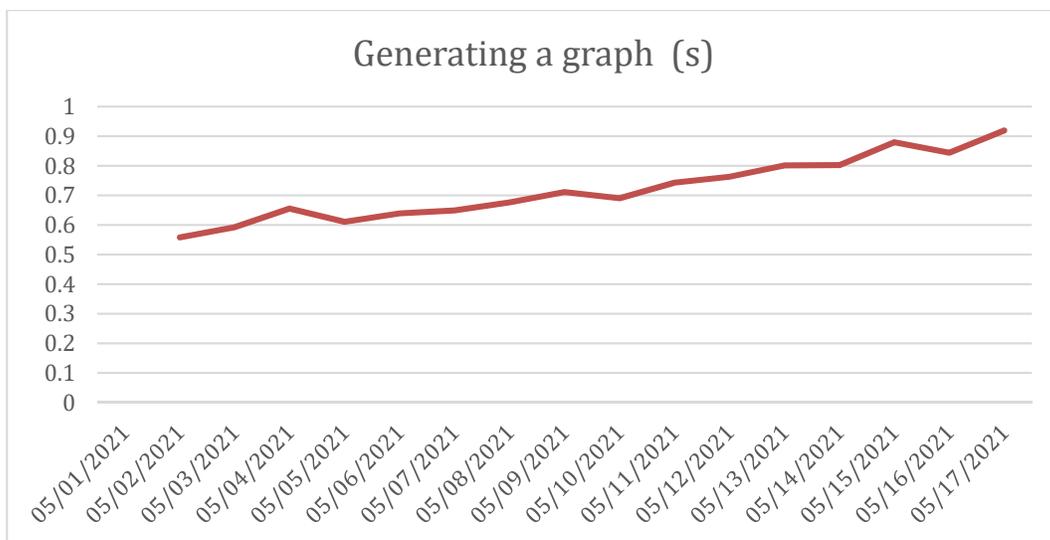

*Figure 9.* The time to generate the graph G
Note. The time to generate graph G ranged from 0.5582 s to 0.9741 s, which means an increase of 0.4159 s.

From the point of view of e-commerce systems, time is a very important parameter. This parameter has also been measured for graph G. By comparing the size parameters of graph G with the time needed to generate that graph, one can see the relationship between these parameters. This relationship is illustrated in Figure 10.

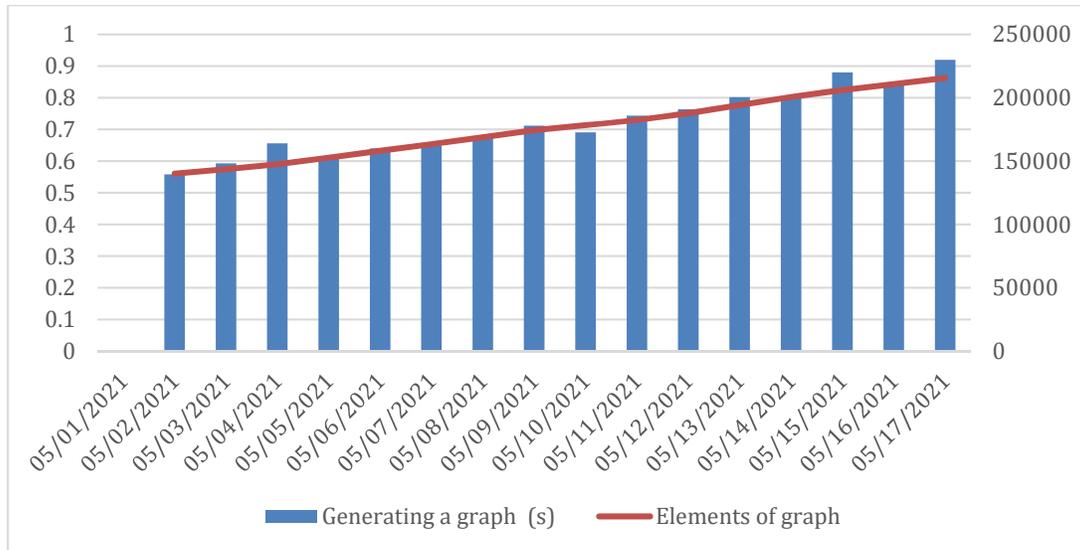

*Figure 10.* The relationship between the number of elements and the time to generate the graph G

Note. As the number of elements of the G graph increases, the time to generate the graph increases.

*The parameters of recommendation systems*

In line with the work's aim consisting of the implementation of the ARS algorithm in an e-commerce solution, a recommendation system functioning in the AM76 IT system was created. This system was tested within May 1-17, 2021. The research results are presented below.

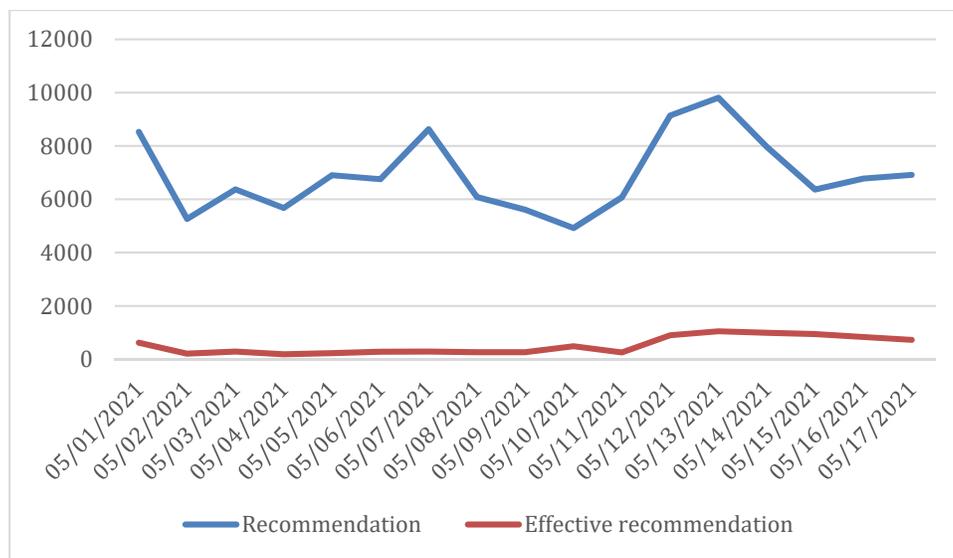

*Figure 11.* Number of recommendations and number of effective recommendations

Note. The number of recommendations in the analysed period ranged from 4926 to 9813, of which 187 to 1052 recommendations per day were effective.

There are various metrics (quality measures) in the area of recommendation systems (Fayyaz et al., 2020). As part of the study, one of them was proposed, based on the ratio of effective recommendations, i.e. those for which the client chose one of the recommended products, to the number of all recommendations. It has been called *effectiveness* and is calculated according to the formula:

$$effectiveness = \frac{the\ number\ of\ effective\ recommendations}{the\ number\ of\ recommendations} *100\% \quad\quad (5)$$

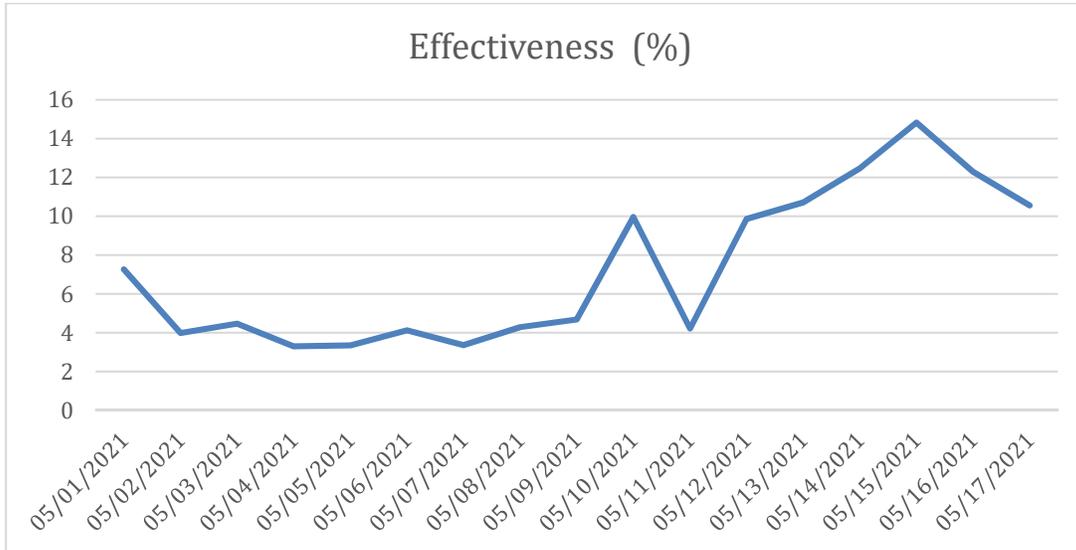

*Figure 12.* Effectiveness of recommendations

Note. The effectiveness in the analysed time period ranged from 3.2% to 14.84%

In order to accurately determine whether the algorithm/recommendation system is effective for this measure, a comparison should be made with other recommendation systems/algorithms, which is not the aim of this study.

Another important parameter in the context of using the ARS algorithm in the e-commerce system is the time of generating recommendations for a given object (product).

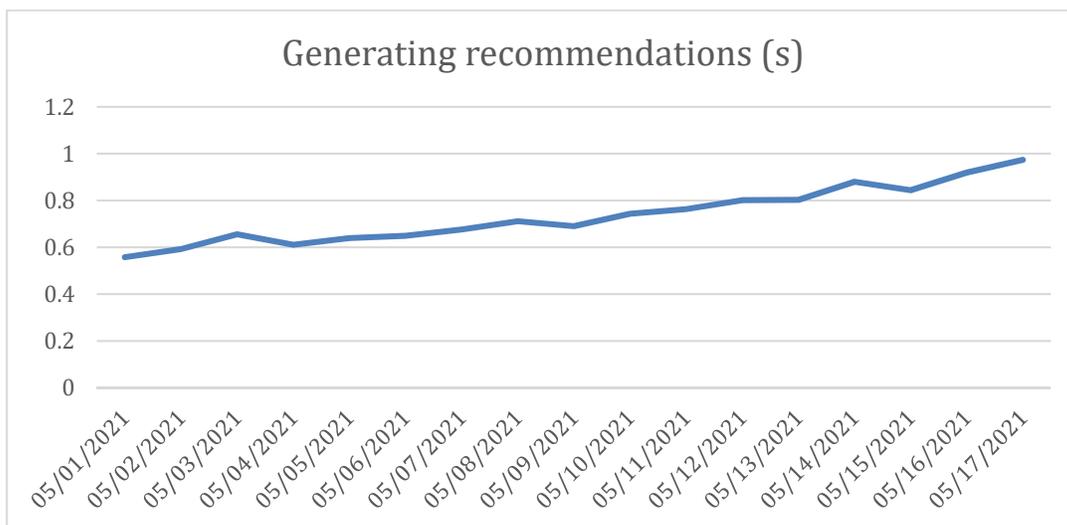

*Figure 13.* The time of generating recommendations for an object (product).

Note. The time of generating recommendations in the examined time period ranged from 0.56 s to 0.97 s.

As shown in Figure 13, this time increases on subsequent days of the system's operation. Based on this and Figure 10, we can hypothesise that this parameter depends on the size of the graph G.

**Discussion**

The ARS algorithm in its basic form is historic in terms of users' actions. This means that the history of activities does not matter for the algorithm to function.Moreover, when it comes to the **G** graph model, all kernel classes have the same meaning. In fact, this translates into the fact that a session built on the basis of website visits has the same importance for the algorithm as a session built on the basis of an order placed and paid by the customer, which seems like it should be differentiated. The purchase of an item should indicate that the item is more attractive and thus should carry more recommendation weight than an item that has only been viewed.

The following modifications can be made to overcome these two drawbacks of the original ARS.

*Modification 1 – adding weights to arcs*

At the stage of preparation of graph **G**, the weights' values should be associated with the arcs coming out of kernels being members of the individual classes. Due to this procedure, the property shall be given that some kernels' classes, such as expert recommendations or purchases, are more valuable than random visits by anonymous users.

In the case of this modification, the mathematical model of the directed graph **G**, called the recommendation session graph, should be considered and supplemented with the function **W** assigning each arc a natural number interpreted as class weight.

*Modification 2 – constructing the user's paths*

In order to eliminate the algorithm's historic agnosticism, an auxiliary variable should be implemented, within the algorithm, in the form of the set **M**, in which information about "viewed" objects shall be stored (nodes of the graph). The first element of this set is the **m** object. The next elements (objects) added to the **M** set are new "visited" objects selected by the user. The steps of the algorithm would then depend not on a single node **m**, but on the set of nodes $M = \{m_1, m_2, m_3, m_4, \ldots, m_n\}$.

**Implications**

The aim presented at the beginning of the study, in the form of the implementation of the ARS algorithm in the e-commerce solution, has been achieved. The recommendation system built this way is fully functional, which is confirmed by the results of the conducted research. The implemented ARS algorithm allows you to increase sales through personalised recommendations matched to the customer's expectations. Well-prepared input data in the form of the graph G, on the basis of which recommendations are then created, gives a chance for a good match to the interests and preferences, due to which customers make purchasing decisions faster and easier. The use of ARS has been shown in e-commerce, but this algorithm can also be used in dedicated websites like thematic catalogues (e.g. accommodation facilities, car market) or sector-specific portals (e.g. board game players, pet lovers).

**Acknowledgments**

Not applicable.

**Funding**

Not applicable.

**Conflict of interests**

No, there are no conflicting interests.